\documentstyle[aps,preprint,epsfig]{revtex}
\tightenlines

\begin{document}
\draft
\title{Lattice Boltzmann Study of Velocity Behaviour 
in Binary Mixtures Under Shear}
\author{Aiguo Xu$^{1}$
\footnote{E-mail address: Aiguo.Xu@ba.infn.it}
 \and G. Gonnella$^{1,2}$}
\address{
$^{1}$ Istituto Nazionale per la Fisica della Materia, Unit\`a di Bari,\\
{\rm and} Dipartimento di Fisica, Universit\`a di Bari, {\rm and}\\
TIRES, Center of Innovative Technologies for Signal Detection\\
and Processing,\\
via Amendola 173, 70126 Bari, Italy\\
$^{2}$ INFN, Sezione di Bari, via Amendola 173, 70126 Bari, Italy }
\maketitle

\begin{abstract}
We apply lattice Boltzmann methods to study the relaxation of the velocity
profile in  binary fluids under shear during spinodal decomposition.
In  simple fluids, when a shear flow is applied on the boundaries of the 
system, the time required to obtain a triangular profile
 is inversely proportional to the viscosity and proportional to
the square of the size of the system. We find that the same behaviour
also occurs for  binary mixtures, for any 
component ratio in the mixture  and independently from
the time when shear flow is switched on during phase
separation.
\end{abstract}

\newpage

\section{Introduction}

Recently, the dynamical behaviour of complex fluid and binary mixtures under
the action of applied flows has been investigated in many theoretical and
experimental works \cite{larson}. In  phase separating  binary mixtures
the morphology and the growth properties of the domains of the two components 
 are greatly affected by the
presence of a flow\cite{Onuki}. This flow is generally imposed from the
boundaries onto the system and takes some time to reach its steady value. If
this time is comparable with other typical time-scales in the system,
transient effects have to be carefully considered. For example, in the case
of an oscillating shear flow, the ratio between the period of the applied
flow and the relaxation time of a triangular steady shear profile is
relevant for the effectiveness of the flow on the bulk of the system. In
spinodal decomposition, if that ratio is small, the growth remains mainly
isotropic and elongated domains can be observed only in the boundary layers 
\cite{XGL2002}.

The estimation of relaxation times of applied flows is generally based on
analogies with simple fluids. However, a careful checking of the laws valid
for simple fluids in cases when mesoscopic structures are present in the
system - e.g. interfaces between domains in spinodal decomposition, cannot
be found in literature. In this paper we evaluate numerically the
relaxational properties of a triangular shear flow applied on phase
separating binary mixtures. At a certain time during phase separation the
top and the bottom boundaries of our system start to move with velocities $w$
and $-w$ respectively and after a while a triangular profile will appear. We
study the evolution of the velocity field in terms of the viscosity and of
the size of the system and  find that the results for a simple fluid also
hold in the case of binary mixtures with interfaces inside.

We simulate the binary mixture system by applying lattice Boltzmann methods
which allow to mimic the behaviour of a system described by Navier-Stokes
and convection-diffusion equations. The lattice Boltzmann algorithm is based
on a collision step and a propagation step occurring on time $\Delta t$.
Only a finite sets of local velocities are allowed. In particular we use a
scheme based on a free-energy approach which has the advantage that the
equilibrium thermodynamics of the system is ``{\it a priori}'' known. The
role of the simulation time step $\Delta t$ will be also considered. With
fast relaxing velocity profiles, the usual choice $\Delta t=1$ is not
appropriate for obtaining smoothly relaxing profiles and smaller $\Delta t$
have to be considered.

In the next section we briefly describe the methods used in the simulations;
section III contains our results and section IV some conclusions.

\section{The model}

We consider a two-dimensional binary fluid with components A and B of number
density $\rho _A$ and $\rho _B$, respectively. Such a system can be modeled
by the following free energy functional,

\begin{equation}
F=\int d{\bf r}[\frac 13\rho \ln \rho +\frac a2\varphi ^2+\frac b4\varphi ^4+%
\frac \kappa 2(\nabla \varphi )^2]\text{, }  \label{eqd1}
\end{equation}
where $\rho =\rho _A+\rho _B$ is the local total density and $\varphi =\rho
_A-\rho _B$ is the local density difference and the order parameter of the
system; the term in $\rho $ gives rise to a positive background pressure and
does not affect the phase behavior. The terms in $\varphi $ correspond to
the usual Ginzburg-Landau free energy typically used in studies of phase
separation\cite{28}. The polynomial terms are related to the bulk properties
of the fluid. The gradient term is related to the interfacial properties.
The parameter $b$ is always positive, while the sign of $a$ distinguishes
between a disordered ($a>0$) and a segregated mixture ($a<0$) where the two
pure phases with $\varphi =\pm \sqrt{-a/b}$ coexist. In this paper we will
consider quenches into the coexistence region with $a<0$ and $b=-a$, so the
equilibrium values for the order parameter are $\varphi =\pm 1$. The initial state in simulations will be random configurations corresponding to the high
temperature disordered phase.

The thermodynamic properties of the fluid follow directly from the free
energy (\ref{eqd1}). The chemical potential difference between the two
fluids is given by 
\begin{equation}
\Delta \mu =\frac{\delta F}{\delta \varphi }=a\varphi +b\varphi ^3-\kappa
\nabla ^2\varphi \text{. }  \label{eqd4}
\end{equation}
The pressure is a tensor $P_{\alpha \beta }$ since interfaces in the fluids
can exert non-isotropic forces\cite{46}. A suitable choice is

\begin{equation}
P_{\alpha \beta }=p_0\delta _{\alpha \beta }+\kappa \partial _\alpha \varphi
\partial _\beta \varphi \text{,}  \label{eqd5}
\end{equation}
where the diagonal part $p_0$ can be calculated using thermodynamics
relations from (\ref{eqd1}):

\begin{eqnarray}
p_0 &=&\rho \frac{\delta F}{\delta n}+\varphi \frac{\delta F}{\delta \varphi 
}-f(\rho ,\varphi )  \nonumber \\
&=&\frac \rho 3+\frac a2\varphi ^2+\frac{3b}4\varphi ^4-\kappa \varphi
(\nabla ^2\varphi )-\frac \kappa 2(\nabla \varphi )^2  \label{eqd6}
\end{eqnarray}
\vspace{3mm}

Our simulations are based on the lattice Boltzmann scheme developed by
Orlandini et al\cite{Or39}. and Swift et al.\cite{Sw40} . In this scheme the
equilibrium properties of the system can be controlled by introducing a free
energy which enters properly into the lattice Boltzmann model. The scheme
used in this paper is based on the D2Q9 lattice: A square lattice is used in
which each site is connected with nearest and next-nearest neighbors. The
horizontal and vertical links have length $\Delta x$, the diagonal links $%
\sqrt{2}\Delta x$, where $\Delta x$ is the space step. Two sets of
distribution functions $f_i({\bf r},t)$ and $g_i({\bf r},t)$ are defined on
each lattice site ${\bf r}$ at each time $t$. Each of them is associated
with a velocity vector ${\bf e}_i$. Defined $\Delta t$ as the simulation
time step, the quantities ${\bf e}_i\Delta t$ are constrained to be lattice
vectors so that $|{\bf e}_i|=\Delta x/\Delta t\equiv c$ for $i=1$, $2$, $3$, 
$4$ and $|{\bf e}_i|=\sqrt{2}c$ for $i=5$, $6$, $7$, $8$. Two functions $f_0(%
{\bf r},t)$ and $g_0({\bf r},t)$, corresponding to the distribution
components that do not propagate (${\bf e}_0=0$), are also taken into
account. The distribution functions evolve during the time step $\Delta t$
according to a single relaxation time Boltzmann equation\cite{42,43}:

\begin{equation}
f_i({\bf r}+{\bf e}_i\Delta t,t+\Delta t)-f_i({\bf r},t)=-\frac 1\tau [f_i(%
{\bf r},t)-f_i^{eq}({\bf r},t)],  \label{lbmeq1}
\end{equation}

\begin{equation}
g_i({\bf r}+{\bf e}_i\Delta t,t+\Delta t)-g_i({\bf r},t)=-\frac 1{\tau
_\varphi }[g_i({\bf r},t)-g_i^{eq}({\bf r},t)],  \label{lbmeq2}
\end{equation}
where $\tau $ and $\tau _\varphi $ are independent relaxation parameters, $%
f_i^{eq}({\bf r},t)$ and $g_i^{eq}({\bf r},t)$ are local equilibrium
distribution functions. Following the standard lattice Boltzmann
prescription, the local equilibrium distribution functions can be expressed
as an expansion at the second order in the velocity ${\bf v}$\cite{40,41}: 
\begin{equation}
\begin{array}{lll}
f_0^{eq} & = & A_0+C_0v^2 \\ 
f_i^{eq} & = & A_I+B_Iv_\alpha e_{i\alpha }+C_Iv^2+D_Iv_\alpha v_\beta
e_{i\alpha }e_{i\beta }+G_{I,\alpha \beta }e_{i\alpha }e_{i\beta } \\ 
&  & i=1,2,3,4, \\ 
&  &  \\ 
f_i^{eq} & = & A_{II}+B_{II}v_\alpha e_{i\alpha }+C_{II}v^2+D_{II}v_\alpha
v_\beta e_{i\alpha }e_{i\beta }+G_{II,\alpha \beta }e_{i\alpha }e_{i\beta }%
\text{,} \\ 
&  & i=5,6,7,8\text{, } \\ 
&  & \text{ }
\end{array}
\label{lbmeq3}
\end{equation}
and similarly for the $g_i^{eq}$, $i=0$, $...$, $8$. The expansion
coefficients $A_0$, $A_I$, $A_{II}$, $B_I$, $\cdots $ are determined by
using the following relations

\begin{equation}
\begin{array}{llll}
\sum_if_i^{eq}({\bf r}\text{, }t)=\rho \text{,} &  &  & \sum_ig_i^{eq}({\bf r%
}\text{, }t)=\varphi \text{,}
\end{array}
\label{lbmeq4}
\end{equation}
\begin{equation}
\begin{array}{llll}
\sum_if_i^{eq}({\bf r}\text{, }t){\bf e}_i=\rho {\bf v}\text{,} &  &  & 
\sum_ig_i^{eq}e_{i\alpha }=\varphi v_\alpha \text{, }
\end{array}
\label{lbmeq5}
\end{equation}
\begin{equation}
\begin{array}{llll}
\sum_if_i^{eq}e_{i\alpha }e_{i\beta }=c^2P_{\alpha \beta }+\rho v_\alpha
v_\beta \text{, } &  &  & \sum_ig_i^{eq}e_{i\alpha }e_{i\beta }=c^2\Gamma
\Delta \mu \delta _{\alpha \beta }+\varphi v_\alpha v_\beta \text{,}
\end{array}
\label{lbmeq6}
\end{equation}
where $P_{\alpha \beta }$ is the pressure tensor, $\Delta \mu $ is the
chemical potential difference between the two fluids and $\Gamma $ is a
coefficient related to the mobility of the fluid. We stress that we are
considering a mixture with two fluids having the same mechanical properties
and, in particular, the same viscosity. The second constraint in Eq. (\ref
{lbmeq5}) expresses the fact that the two fluids have the same velocity.

A suitable choice of the coefficients in the expansions (\ref{lbmeq3}) is
shown in Ref.\cite{XGL2002}. Such a lattice Boltzmann scheme simulates at
second order in $\Delta t$ the continuity, the quasi-incompressible
Navier-Stokes and the convection-diffusion equations with the kinematic
viscosity $\nu $ and the macroscopic mobility $\Theta $ given by\cite
{40,41,Chapman48} 
\begin{equation}
\begin{array}{llll}
\nu =\Delta t\frac{c^2}3(\tau -\frac 12)\text{, } &  &  & \Theta =\Gamma
\Delta tc^2(\tau _\varphi -\frac 12)\text{. }
\end{array}
\label{lbmeq7}
\end{equation}

The shear flow can be imposed by introducing boundary walls on the top and
bottom rows of lattice sites. The velocities of the upper and lower walls are
along the horizontal direction and their values are $w$ and $-w$,
respectively, where $w=\gamma (L-1)/2$ and $\gamma $ is the shear rate
imposed on the system. The bounce-back rule\cite{bb1,bb2} is adopted for the
distribution functions normal to the boundary walls. In order to preserve
correctly mass conservation, a further constraint, related to the
distribution functions at the previous time step $(t-\Delta t)$, is used.
Details of the scheme are given in Ref. \cite{La36}. 

Finally, we observe that if we set $a=b=\kappa =0$ in the free energy
functional (\ref{eqd1}), the present lattice Boltzmann methods can be used
to simulate simple fluids.

\section{Results}

We numerically check the relaxation behaviors of the horizontal velocity
profile for symmetric and asymmetric binary fluids. We consider two cases:
(i) switch on the shear from the beginning of the phase separation and (ii)
switch on the shear during the phase separating process and after the
interfaces between the two fluids have formed. We denote the time at which
the shear is switched on as $t_{on}$. We focus on the effects of viscosity,
so we vary $\tau $ in the simulations. The other parameters, if not
differently stated, are fixed at the values: $-a=b=1.252\times 10^{-4}$, $%
\kappa =8\times 10^{-5}$, $\Gamma =40$, $\Delta x=1$, $\Delta t=0.2$, $%
\gamma =0.005 $. Similar results are obtained for other sets of parameters.

Figure 1 shows the relaxation of the horizontal velocity profile for a
symmetric binary fluid with $\tau =2$, $L=64$, $t_{on}=5000$. The four lines
correspond to $t=5020$, $t=5040$, $t=5060$, $t=5120$. From this figure we
can see how the shear flow comes into the bulk of the system. When $t>5120$
the velocity profile shows an almost linear behavior with $y$. Before
switching on the shear the domains grow isotropically. After $t_{on}$
anisotropic behavior come into the bulk of the system with time. Figure 2
shows the configuration at the time $t=5120$. Domains separated by well
formed interfaces can be observed to incline to the flow direction with
time. For the case $t_{on}=0$ we observed the same behavior for the velocity
profile, while the interfaces have not reached their equilibrium shape. That
means the existence of the interfaces does not influence the shear effects
coming into the system. For the asymmetric case, we studied the binary fluid
with the ratio $80:20$. After switching on the shear the velocity profile
shows the same behavior.

To understand better the time evolution of the velocity profile of binary
mixtures, we take the simple fluid as a reference. We consider the Newtonian
viscous flow between two infinite plates with a distance of $L$. We use $y$
to denote the coordinate in the vertical direction and $y=0$ in the middle
of the system. The velocities of the upper and the lower plates are $w$ and $%
-w$, respectively . The shear rate imposed on the system is $\gamma $. 
The motion equation
is 
\begin{equation}
\rho \frac{\partial u}{\partial t}=\eta \frac{\partial ^2u}{\partial y^2}%
\text{ }  \label{eqm4}
\end{equation}
The velocity profile can be obtained by standard methods\cite{velpro} and is
given by 
\begin{equation}
u=\gamma y-\sum_n(-1)^{n+1}\frac{\gamma L}{n\pi }\exp (-\frac{4n^2\pi ^2\nu 
}{L^2}t)\sin (\frac{2n\pi }Ly)\text{.}  \label{eqm17}
\end{equation}
When $t$ is large enough, the modes with $n\geq 2$ can be neglected, which is 
confirmed by simulations.  
So we
can define a relaxation time for the velocity profile in the following way,

\begin{equation}
T_R=\frac{L^2}{4\pi ^2\nu }=\frac{3L^2}{2\pi ^2\Delta tc^2(2\tau -1)}\text{.}
\label{simeq1}
\end{equation}
It is interesting to see if or not such a definition also works for binary
fluids. To numerically check the relation between $T_R$ and $\nu $ (or $\tau 
$) or $L$, we calculate $T_R$ in the following way,

\begin{equation}
T_R=-\frac t{\ln (\Xi )}\text{, }  \label{simeq3}
\end{equation}
where 
\begin{equation}
\Xi =\frac{\pi (\gamma y-u)}{\gamma L\sin (\frac{2\pi }Ly)}\text{. }
\label{simeq4}
\end{equation}
In order to calculate $T_R$, we use a time $t$ at which the velocity profile
has been almost linear, so that terms with $n\geq 2$ in Eq.(\ref{eqm17}) can
be neglected.

Figure 3 shows the simulation results for $T_R$ as a function of $\nu $,
where $L=128$, $\Delta t=0.2$, the vertical axis is for $1/T_R$ and the
horizontal axis is for $\nu $ so that we expect a linear behavior. The
dotted line with points correspond to the simulation results and the solid
line corresponds to the expected value from the definition (\ref{simeq1}).
The simulation results confirm the existence of the exponential behavior in
the relaxation process of the velocity profile and the validity of the
definition of $T_R$.

Figure 4 shows $T_R$ as a function of $L$, where the vertical axis is for $%
T_R$ and the horizontal axis is for $L^2$ . The dotted lines with symbols
are simulation results for the cases of $\nu =4.17$, $\nu =5.83$ and $\nu
=7.5$. The solid lines are the expected values from the definition (\ref{simeq1}%
). The four points in each case correspond to $L=64$, $128$, $256$, and $512$%
. The simulation step in this figure is $\Delta t=0.2$. We find the expected
linear behavior between $T_R$ and $L^2$.

There is a bending tendency in the simulation results for $1/T_R$ in Fig.3,
so that the simulation results deviate from the expected values. If we
continue the simulation up to a much higher viscosity region, we will find a
more pronounced bending behavior. We emphasize that this is an artificial
phenomenon, which can be mainly attributed to the finite size of the
simulation step $\Delta t$. If the viscosity is larger the velocity profile
relaxes more quickly and we should use a smaller time step $\Delta t$ for
the simulation. When the viscosity is very high and $\Delta t$ is not small
enough we can not observe smooth velocity profiles in the simulation. To
confirm this numerical analysis we show two simulation results in Fig.5,
where the definition value of $1/T_R$ is also shown to guide the eyes.
Compared with the case of $\Delta t=0.2$, when we use $\Delta t=0.1$ the
linear behavior region is almost doubled.

Finally, it is interesting to compare the relaxation behaviors of binary and
simple fluids. Fig.6 shows the simulation results of the velocity profiles
for the case with $L=256$, $\tau =2$ ($\nu =2.5$), $t-t_{on}=600$. The solid
line corresponds to Eq. (\ref{eqm17}) with $n=1$. From this figure the
following remarks are evident: (i) the relaxation process of the binary
mixtures follows the same behavior as that of simple fluids; (ii)for binary
mixtures the relaxation behavior of the velocity profile is independent of
the component ratio and the time when shear is switched on, which means that
the forming of the interfaces does not evidently influence the shear effects 
coming into the system.

\section{Conclusions}

In this paper we study the relaxation behavior of the velocity profile of
binary mixtures under steady shear with lattice Boltzmann methods. Following
the simple Newtonian viscous flow, we define a relaxation time of the
velocity profile $T_R$ whose value is inversely proportional to the
viscosity and proportional to the square of the size of the system. The
simulation results show that the shear effects come into the system in the
same way for binary mixtures and simple fluids, which confirm the validity
of the definition of $T_R$ in this and previous studies\cite{XGL2002}. For
binary mixtures, the shear behavior is independent of the component ratio
and the time at which the shear is switched on. The presence of interfaces
between the two fluids has negligible influence on the relaxation process of
the velocity profile.

\newpage 

\begin{figure}[tbp]
\centerline{\epsfig{file=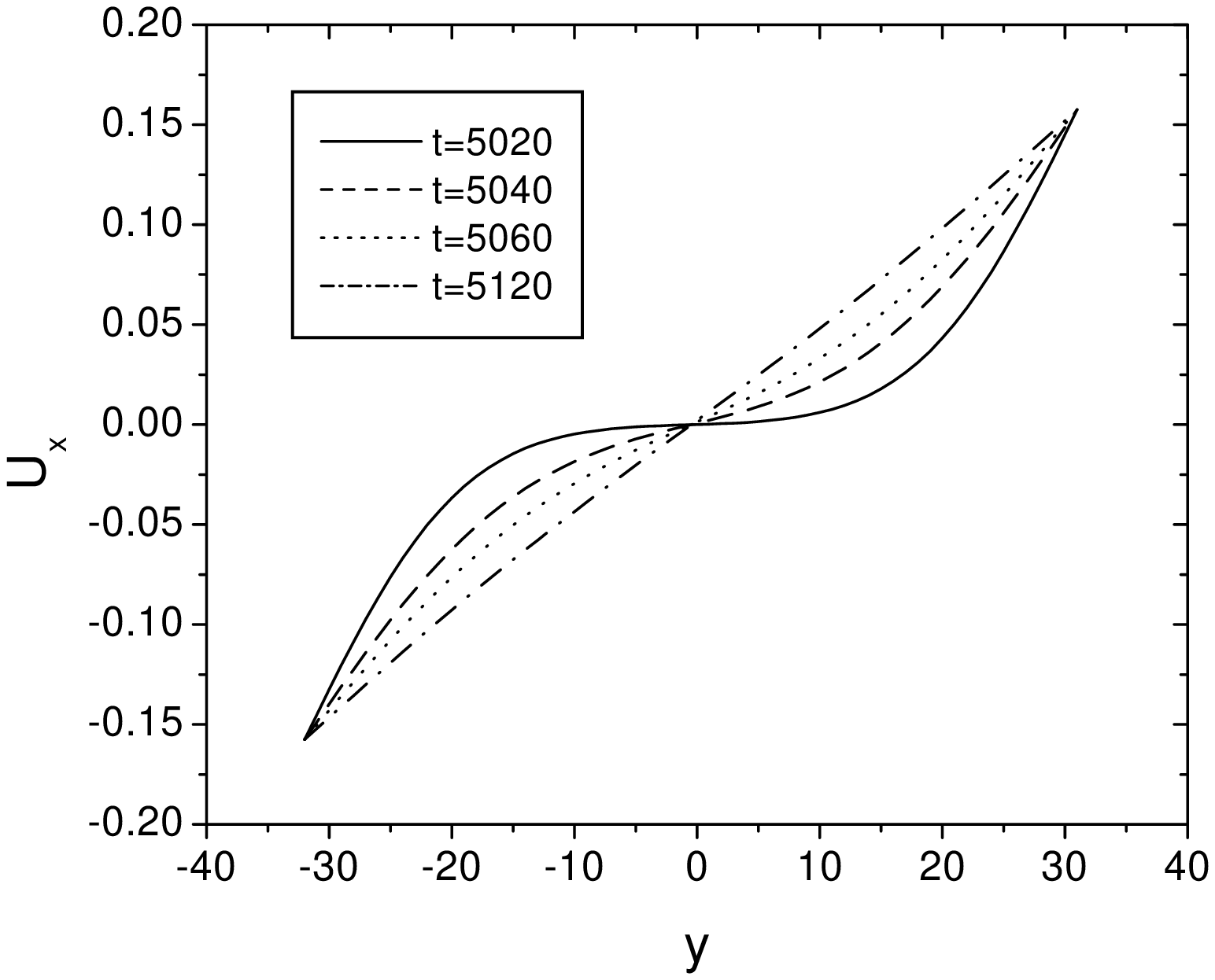,bbllx=36 pt,bblly=371 pt,
bburx=526 pt,bbury=753 pt,
width=0.8\textwidth,clip=}}

\caption{ Relaxation process of the horizontal velocity profile for the
symmetric binary fluids, where $\tau =2$, $L=64$, $t_{on}=5000$. The
corresponding times are shown in the inset. }
\label{fig_1}
\end{figure}

\newpage 
\begin{figure}[tbp]
\centerline{\epsfig{file=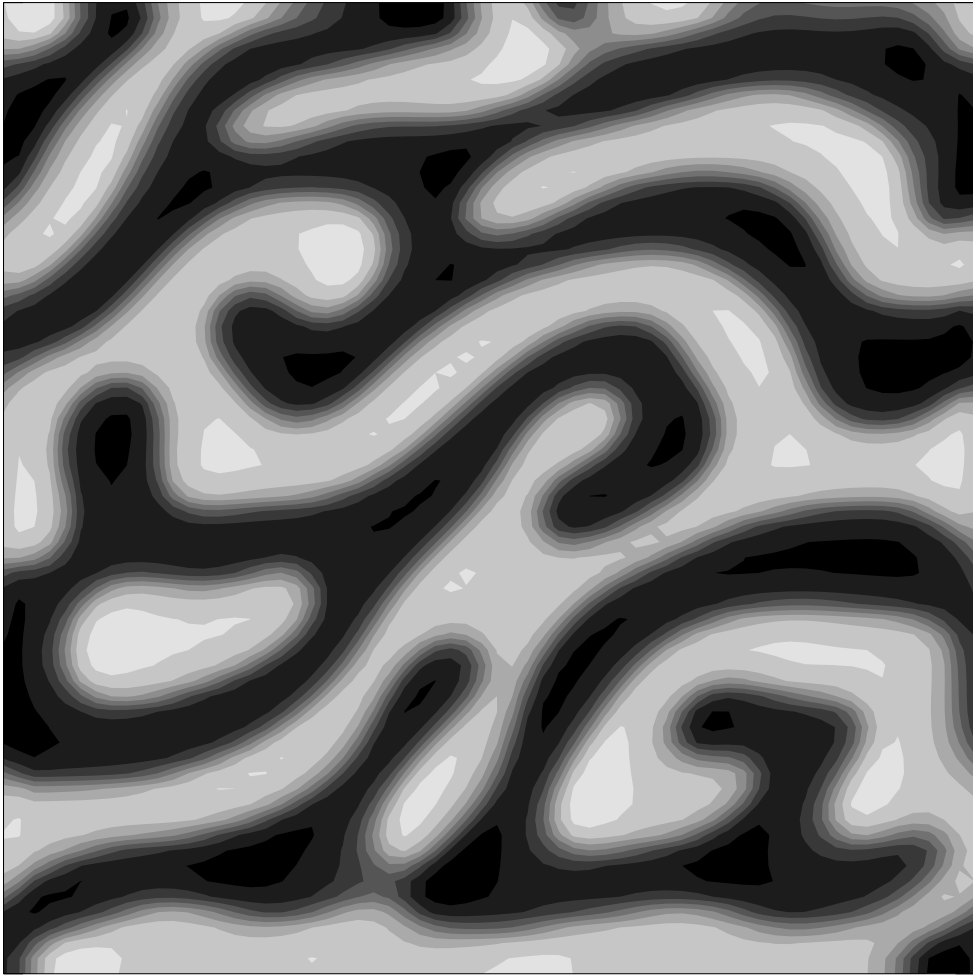,bbllx=48 pt,bblly=357 pt,
bburx=511 pt,bbury=819 pt,
width=0.8\textwidth,clip=}}

\caption{ The configuration of the field $\varphi$ at the time $t=5120$, the
parameters are the same as in Fig.1. From dark to white, the value of $%
\varphi$ varies in the scope $-1.0 \leq \varphi \leq 1.0$. }
\label{fig_2}
\end{figure}

\newpage 
\begin{figure}[tbp]
\centerline{\epsfig{file=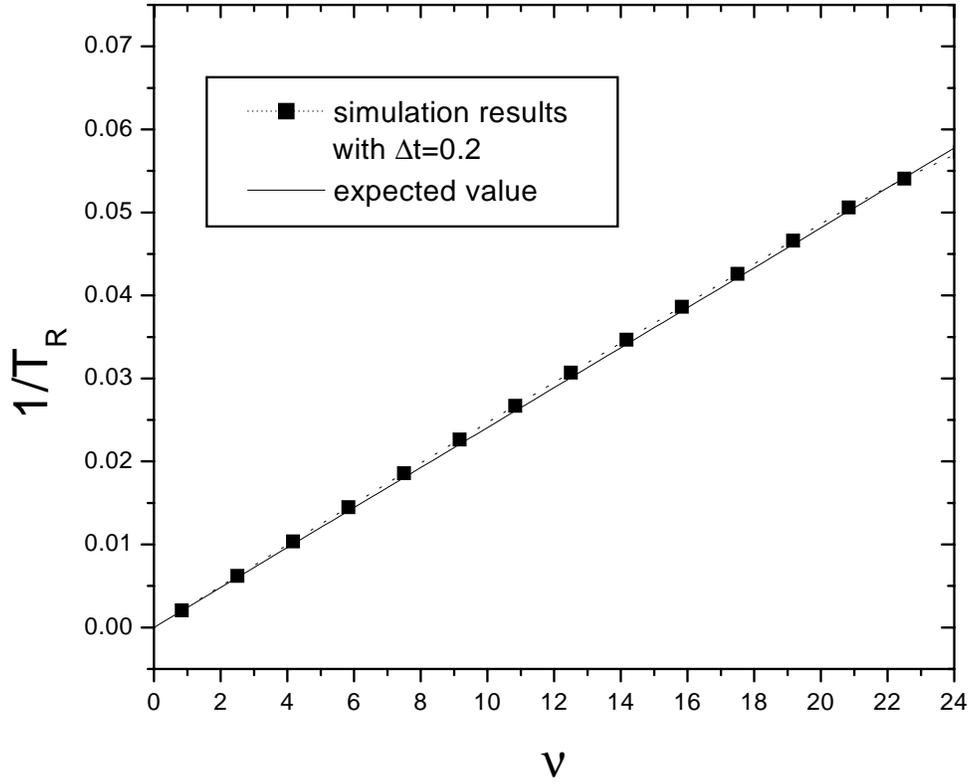,bbllx=53 pt,bblly=346 pt,
bburx=516 pt,bbury=738 pt,
width=0.8\textwidth,clip=}}

\caption{$1/T_R$ as a function of $\nu $. The lattice size $L=128$ is used in 
the simulation.
 The dotted line with symbols correspond to the simulation results and the 
solid line shows the expected values from the definition (\ref{simeq1}). }
\label{fig_3}
\end{figure}

\newpage 
\begin{figure}[tbp]
\centerline{\epsfig{file=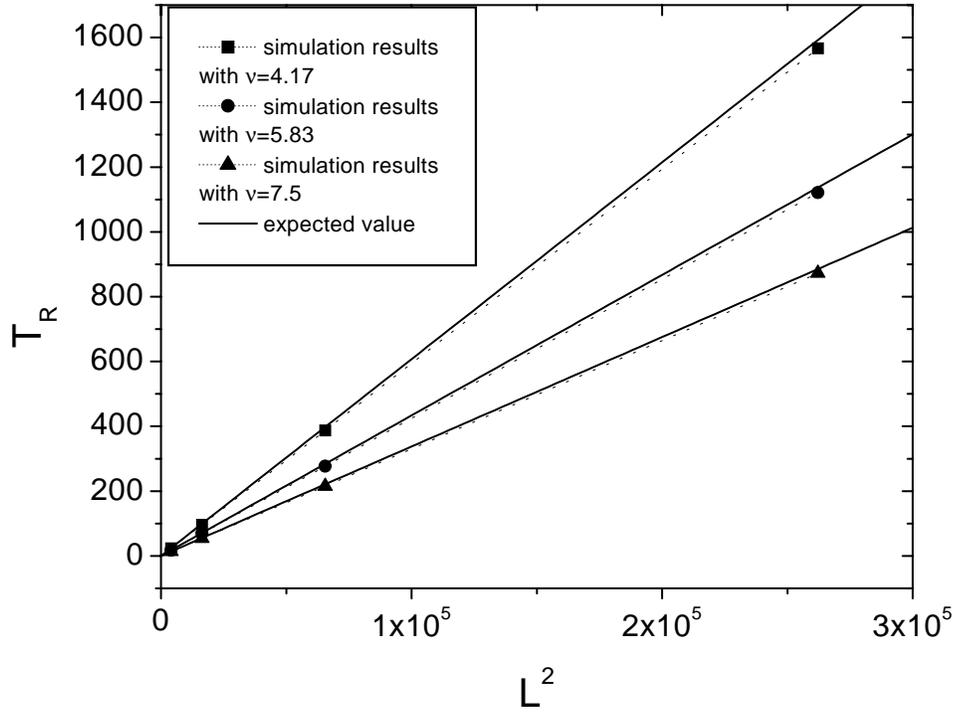,bbllx=43 pt,bblly=355 pt,
bburx=525 pt,bbury=751 pt,
width=0.8\textwidth,clip=}}

\caption{$T_R$ as a function of $L^2$. The dotted lines with symbols show 
the simulation results and the solid lines show the expected values
from the definition (\ref{simeq1}). The four points in each case correspond
to cases of $L=64$, $128$, $256$, and $512$. The simulation step in this figure
is $\Delta t=0.2$. }
\label{fig_4}
\end{figure}

\newpage 
\begin{figure}[tbp]
\centerline{\epsfig{file=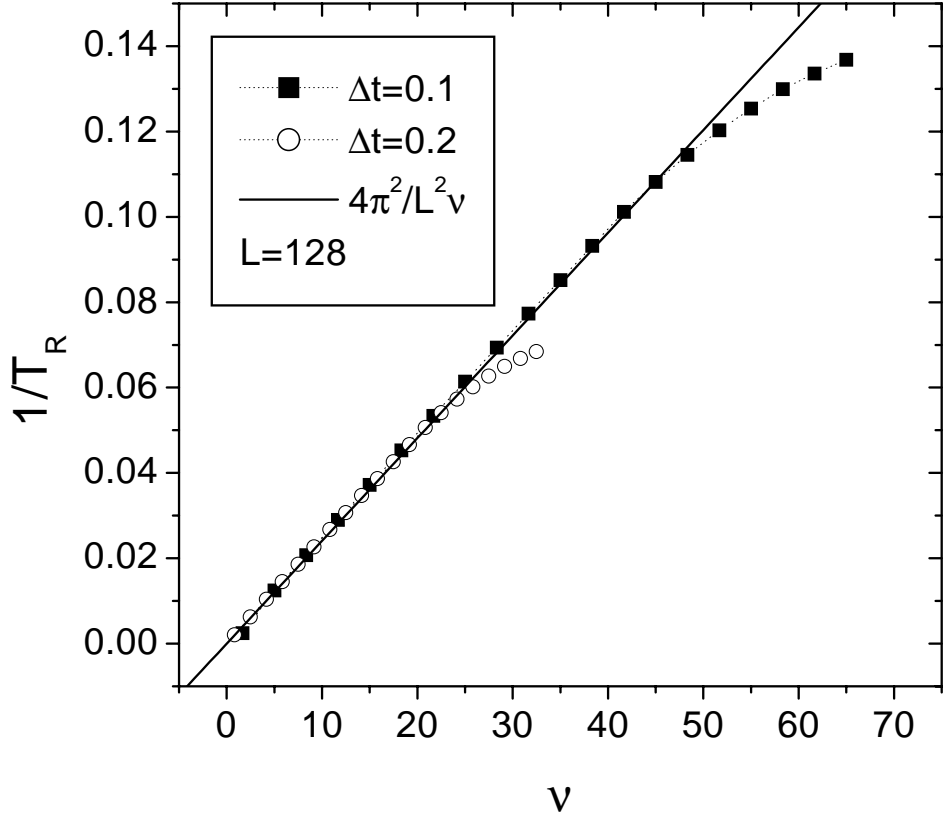,bbllx=53 pt,bblly=322 pt,
bburx=519 pt,bbury=747 pt,
width=0.8\textwidth,clip=}}

\caption{ Two simulation results for $1/T_R$ as a function of $\nu$, where
the expected value of $1/T_R$ is also shown to guide the eyes. The lattice
size and time steps used in the simulations are shown in the inset. }
\label{fig_5}
\end{figure}

\newpage 
\begin{figure}[tbp]
\centerline{\epsfig{file=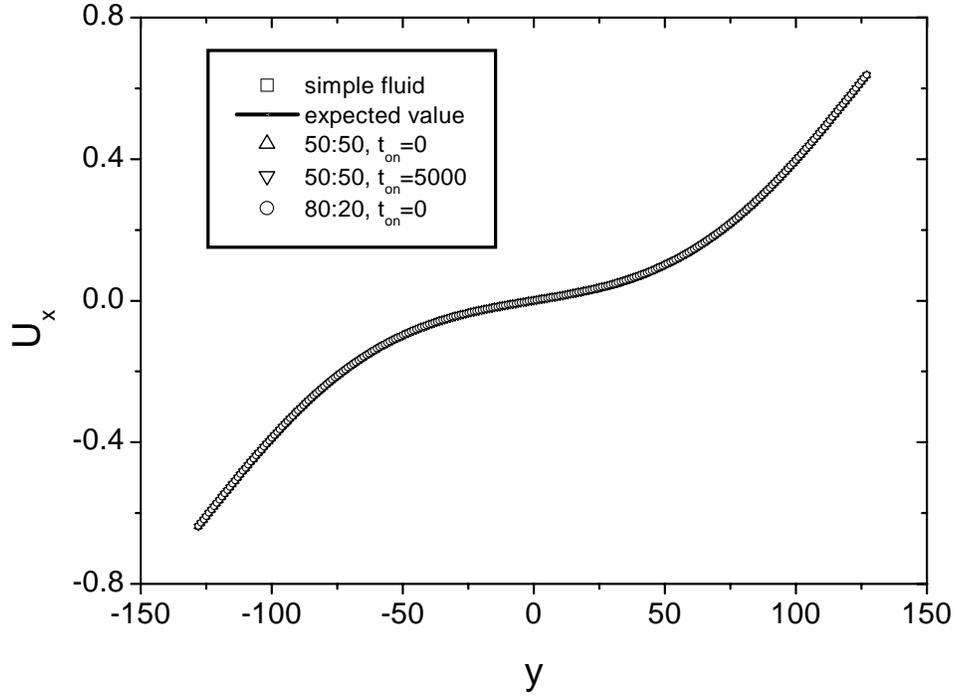,bbllx=99 pt,bblly=474 pt,
bburx=432 pt,bbury=734 pt,
width=0.8\textwidth,clip=}}

\caption{ Simulation and expected results of the velocity profiles, where $L=256$, $\tau =2$ ($\nu =2.5$), $t-t_{on}=600$. The solid line shows the expected 
values from Eq. (\ref{eqm17}) with $n=1$.
The symbols show the simulation results. The type of the fluid and the value of $t_{on}$ are shown in the inset. }
\label{fig_6}
\end{figure}

\end{document}